# scDEFT: A deep learning framework for drug-effect prediction and counterfactual reasoning

**Murthy Devarakonda, PhD**

IEEE/AMIA Fellow, ACM Distinguished Member

Causal Data AI (causaldata.ai)

mvd@acm.org

## Abstract

Longitudinal single-cell atlases now capture matched pre- and post-treatment states from responders and non-responders, presenting an opportunity to mechanistically explain why two patients on the same drug diverge. We introduce scDEFT (single-cell Drug EFfect Transducer), which treats a drug as a conditioning operator on cell representations, enabling prediction and explanation. In scDEFT, feature-wise linear modulation produces drug-conditioned cell latents, learned under abundant per-cell supervision and then frozen. Two independent heads aggregate those latents over shared transcriptional neighborhoods to predict drug-induced state change and responder status. A backward stage ranks the latent dimensions by how strongly they separate responders from non-responders and maps them to genes under a cell-composition control. On a harmonized inflammatory bowel disease atlas of 1.16 million cells, three cohorts and two drug classes, scDEFT predicts state change at 45% of the baseline-to-reproducibility-ceiling headroom and stratifies responders before treatment at AUROC 0.70, where standard predictors remain at chance. These predictions and the drivers behind them support target and co-target nomination, patient stratification, and counterfactual prediction of unseen drug–cohort effects.

## Introduction

An important question in treatment biology is why two patients receiving the same drug follow different trajectories. Single-cell profiling of matched pre- and post-treatment biopsies, now available for several diseases and drug classes, makes this question approachable for the first time using AI. The same donor is observed before and after drug treatment, and donors are labeled responders (R) or non-responders (NR) by an independent clinical endpoint. Inflammatory bowel disease (IBD) is an ideal setting, because such longitudinal cohorts exist, roughly half of patients fail to respond to anti-tumor-necrosis-factor (anti-TNF) therapy [12,13], and the difference is not explained by any single biomarker.

Existing single-cell perturbation methods [1,7,8,9,16], often using latent representations from single-cell foundation models, were developed to predict the post-state from a pre-state, but recent benchmarks [6,17,18] showed that such models frequently fail to beat simple linear baselines on held-out perturbations. Compositional methods [5] capture one consequence of state dynamics — cells entering or leaving populations — but not the dynamics themselves. Neither is designed to contrast responders with non-responders or to attribute that contrast to interpretable genes. What is missing is a mechanistic approach whose output is the difference between the two journeys, expressed in gene programs.

We designed scDEFT (single-cell Drug EFfect Transducer), a multi-stage deep-learning model and model-analyzer, to produce that insight (Figure 1a). scDEFT treats drug action as a transition operator on scRNA-seq data from human gut biopsies and decomposes the analysis into a forward learning stage and a backward attribution stage. In the training stage a drug-conditioned representation is trained for each pre-treatment cell; prediction heads aggregate those latents by transcriptional neighborhood [5] to predict both neighborhood state change and responder status. The aggregation helps to extract signal from otherwise noisy data.

The attribution stage then traces the most predictive neighborhoods and their latents back into the transcriptome, ranking the dimensions that separate eventual responders from non-responders and attributing them to genes and programs. The programs and co-targets identified are prioritized hypotheses for experiment.

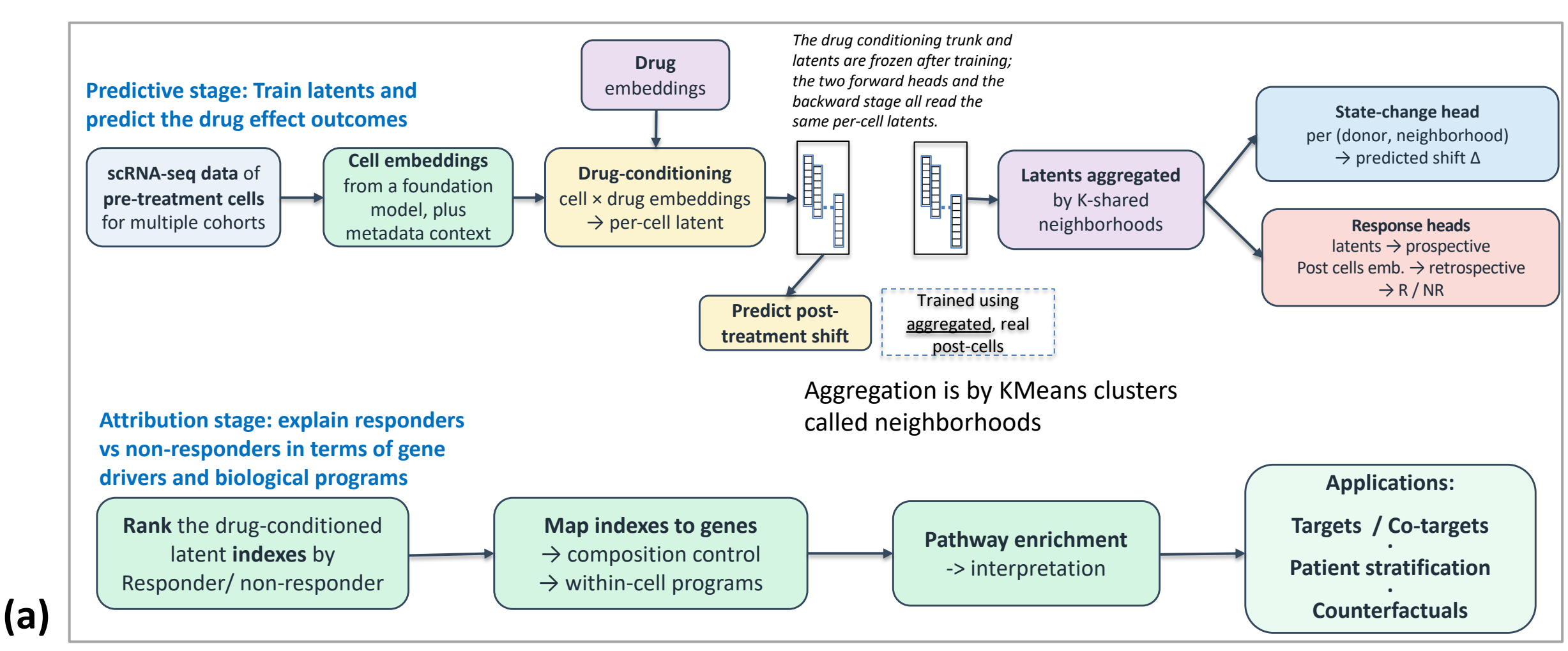

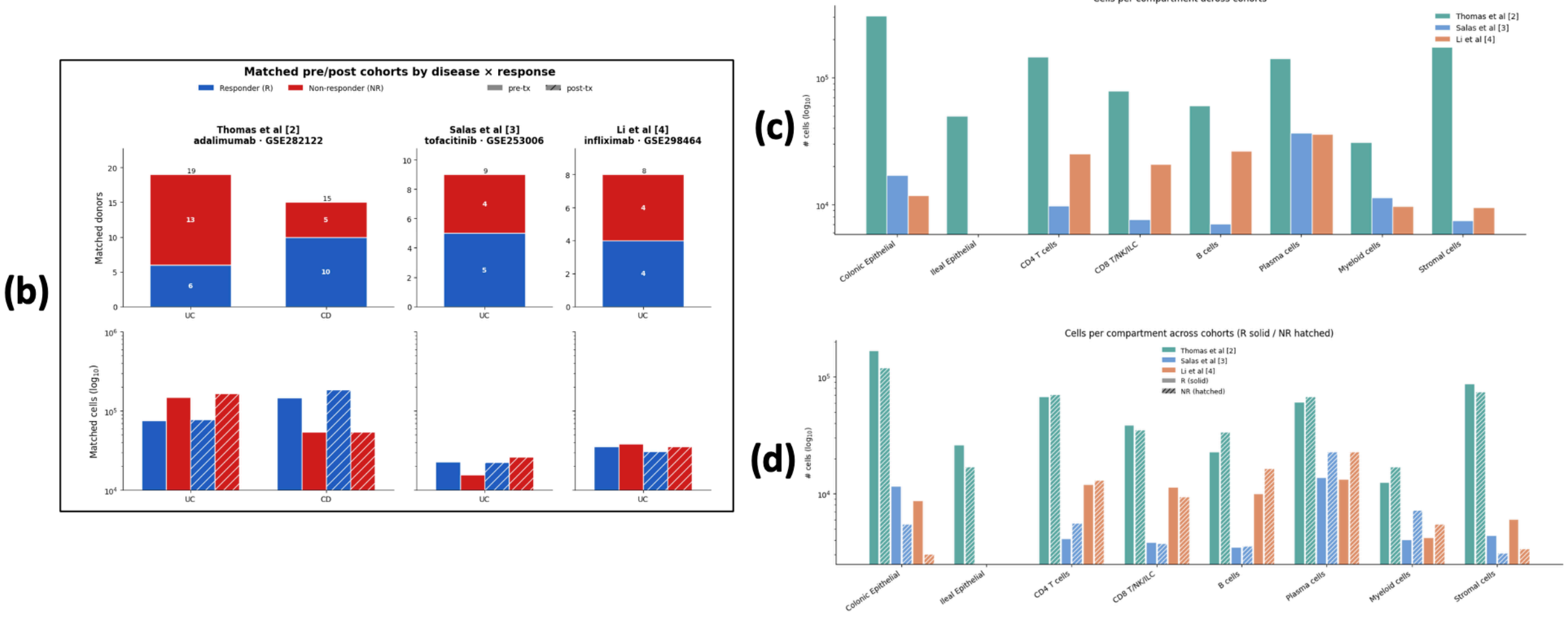

**Figure 1 | scDEFT framework and the harmonized IBD atlas.** *(a) scDEFT predictive stage where cells and a drug are embedded and combined into a per-cell drug-conditioned latent, supervised by a neighborhood-aggregated post-treatment target; the representation is then frozen and read by two decoupled heads: neighborhood state change (Δ) and responder/non-responder prediction, prospective or retrospective. The attribution stage where the same frozen dimensions are ranked by state change and R/NR separation, mapped to genes under a composition control, and named by enrichment, supporting applications of the method. (b) Donors with matched pre- and post-treatment samples and a response label, by cohort, disease and outcome (total 51 donors; responders blue, non-responders red), with matched cell counts below ($log_{10}$; pre-treatment solid, post-treatment hatched). (c) Cells per harmonized compartment ($log_{10}$), colored by cohort. (d) As in (c), split into responders (solid) and non-responders (hashed).*

# Results

## scDEFT learns transcriptomic drug conditioning

scDEFT works with a longitudinal single-cell atlas containing paired pre- and post-treatment samples per donor and a binary response label. The post-treatment cells serve only to define what the model is trained against, i.e., the observed state change, while only the pre-treatment cells are input to the model. Cells are embedded with Geneformer-V2 [1], and drugs with ChemBERTa [26] for small molecules or ESM3 [27] for antibodies (Methods).

The framework then operates in two stages (Figure 1a). The forward stage learns a drug-conditioned cell representation of pre-treatment cells and trains separate heads for neighborhood state-change and response prediction. The attribution stage ranks the latent dimensions that most separate responders from non-responders and attributes them to genes and programs by their correlation with measured expression. scDEFT is deliberately decoupled: the representation is learned from the abundant per-cell state signal, while each downstream task reads it through its own lightweight predictor, so distinct biological questions are not forced into one optimization objective. Neighborhoods serve both as the latent training target and as the aggregation unit inside each head.

We applied scDEFT to a harmonized IBD atlas of 1.16M cells from three published longitudinal cohorts (Figure 1b) spanning two drug classes: the TAURUS anti-TNF (adalimumab) atlas [2], an infliximab cohort [4] and a tofacitinib (JAK-inhibitor) cohort [3] (Figures 1c and 1d). Fifty-one donors had paired samples and response labels, giving 518,437 pre-treatment cells; a shared scaffold of $K = 80$ transcriptional neighborhoods was defined across all cohorts and all cells, so every donor is described in the same coordinate system.

## Drug-conditioning model

### *FiLM layer + Layer Norm & FFN Backbone*

The drug effect on each cell is modeled by conditioning that cell's embedding on the drug through Feature-wise Linear Modulation (FiLM) [20], in which a drug embedding d is mapped to a scale $\gamma(d)$ and a shift $\beta(d)$ that modulate the cell representation e, giving $h = \gamma(d) \odot e + \beta(d)$ (Figure 2; Methods). FiLM keeps the cell representation as the substrate and lets the drug act only as a scale-and-shift, so the same baseline cell is represented differently under different drugs with few added parameters, and any unseen drug with a molecular embedding can condition the same cells, which makes counterfactual queries possible.

The backbone is trained by predicting the change in the representation of each pre-treatment cell and comparing the change with the observed one. No such target exists directly — a cell measured before treatment is never re-measured after it — so scDEFT supplies it through transcriptional neighborhoods.

### *Transcriptional neighborhoods as the unit of biological response*

Treatment-induced state changes are subtle, consistent, and localized to a few cell populations. Annotation-defined cell types conflate two mechanistically distinct changes, a shift in cell-type composition and a change of state within a population, while donor- or sample-level aggregation is too coarse. The transcriptional neighborhood, explored earlier for differential abundance in Milo [5], is the grain that is fine enough to localize a program and coarse enough to be reproducible. Because the scaffold is defined once across all cohorts and then fixed, a neighborhood index denotes the same transcriptional state in every donor (Methods).

## *Training the backbone and the latents*

The entire backbone — the FiLM scale/shift network, the LayerNorm and the FFN — is trained on a single objective: per-cell state-change prediction against the neighborhood-aggregated observed shift (Figure 2; Methods). Once that task converges the per-cell latents are frozen, carrying drug-conditioned state signal learned purely from abundant per-cell supervision, and every downstream application reads this fixed representation.

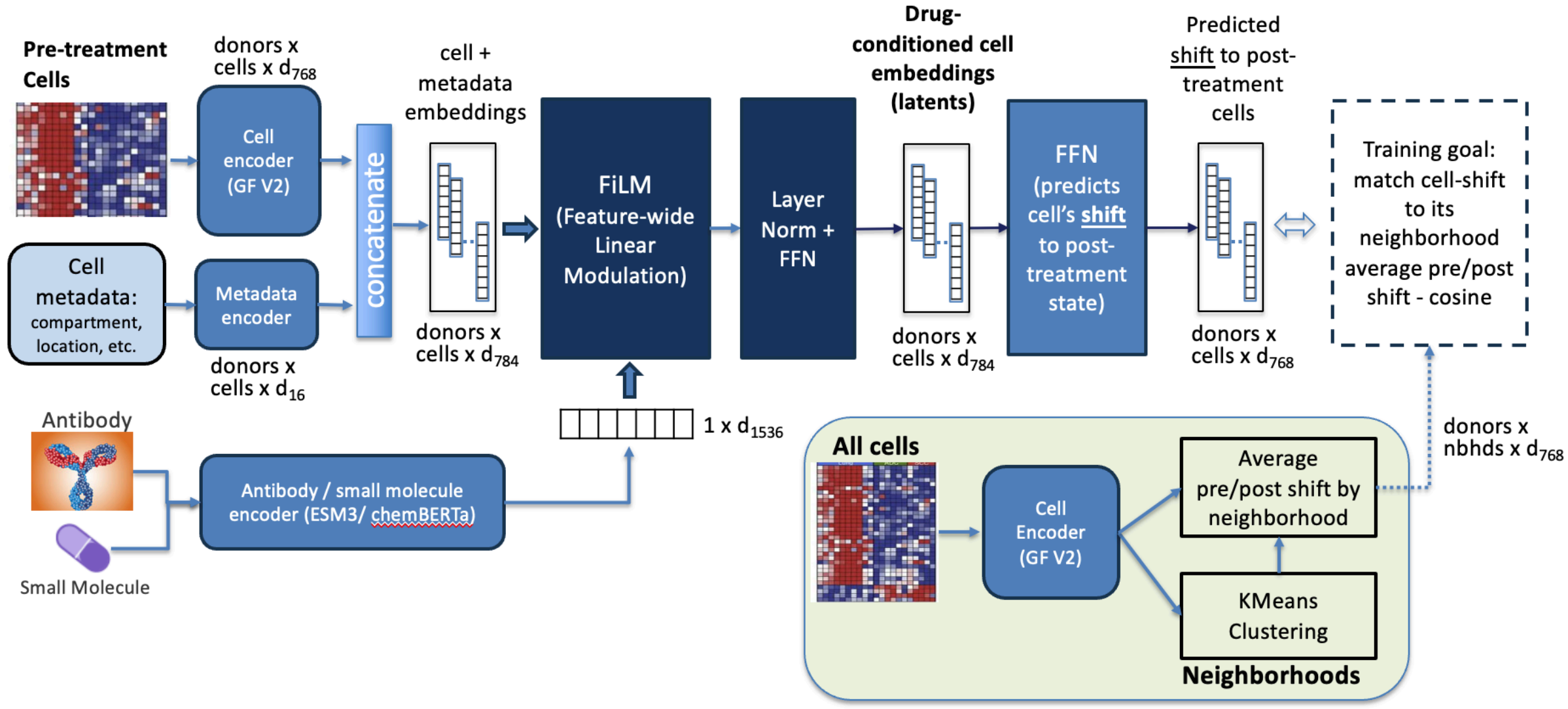

**Figure 2 | Drug conditioning of the cell representation.** *A cell embedding, concatenated with a metadata context vector, is modulated by a drug embedding through Feature-wise Linear Modulation (FiLM), followed by layer normalization and a feed-forward projection. A lightweight head predicts each cell's shift $\hat{\Delta}$ toward the post-treatment state; the whole backbone is trained on this single objective under a per-cell cosine loss against the neighborhood-aggregated observed shift.*

## *Downstream task 1 — predicting drug-induced state change of a donor*

The per-cell objective trains the representation where supervision is abundant but is too noisy for donor-level prediction. A separate donor-level head reads the frozen latents and predicts the state change each (mean-pooled) neighborhood undergoes under the drug effect. This separates the level at which the representation is learned from the level at which it is used.

Predictions are scored on a standardized scale bounded by two references computed on the same folds: a shared-drug baseline, the average shift of training donors in a neighborhood (0.15), and a ceiling given by the split-half reproducibility of the observed shift (0.72). Performance is reported as the gain above baseline as a fraction of that headroom (Methods).

A per-cell state model reached a gain of +0.082 over the baseline. Aggregating to the neighborhood with an attention-based multiple-instance learning head [14] (Figure 3a) more than triples this, to +0.273, or 45% of the reproducibility headroom (Figure 3b). Therefore, neighborhoods, not individual cells, carry the reproducible donor-specific drug-effect signal.

## *Downstream task 2 — predicting response*

The second forward head predicts whether a donor will respond, in two pipelines that differ in the information available. The prospective pipeline (Figure 3c) uses pre-treatment cells only — the harder and more valuable task. Frozen latents are mean-pooled per neighborhood, averaged per donor, reduced to 30 principal components and classified by L2-regularized logistic regression; the reduction is necessitated by the small number of donors (51). Under donor-grouped ten-fold cross-validation this reaches an AUROC of 0.70 [95% CI 0.56–0.84], significantly outperforming a mean-pool ablation of the same latents that removes the neighborhood step (Figure 3d).

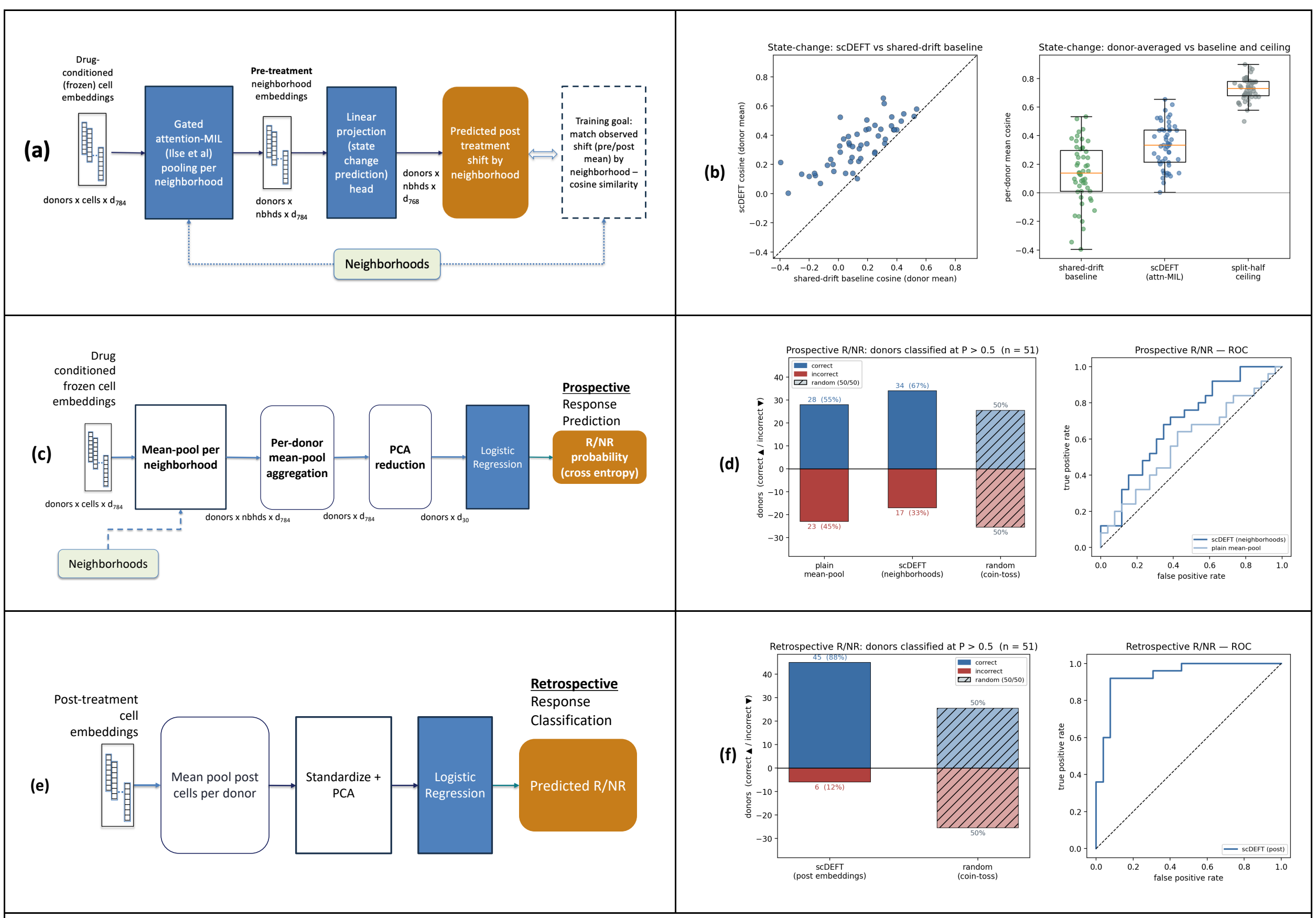

**Figure 3 | State-change and response prediction and evaluation***. (a) The state-change head applies gated attention-based multiple-instance pooling over neighborhood latents, followed by a linear projection to predict shift Δ. (b) State-change results. Left: per-donor mean cosine with the observed shift on held-out donors, scDEFT against the shared-drug-effect baseline; points above the diagonal are donors scDEFT predicts better. Right: the same donor-averaged cosine for the baseline, scDEFT and the split-half ceiling — scDEFT recovers ~45% of the headroom between them. (c) Prospective response pipeline: frozen latents of pre-treatment cells are mean-pooled within neighborhoods for each donor, averaged per donor, reduced to (30) principal components and classified. (d) Prospective results, n = 51 donors. Left: donors predicted correctly or incorrectly at P > 0.5, for plain per-donor mean-pooling, scDEFT with neighborhood aggregation, and a 50/50 reference; aggregation lifts correct calls from 28 (55%) to 34 (67%). Right: ROC for the same two models, scDEFT AUROC is 0.70. (e) Retrospective pipeline: post-treatment cell embeddings are mean-pooled to a single vector per donor, reduced to (30) principal components and classified by logistic regression. (f) Retrospective results, n = 51 donors: 45 (88%) classified correctly, AUROC 0.93, under donor-grouped ten-fold cross-validation.*

The retrospective pipeline (Figure 3e) classifies response from post-treatment cells, mean-pooled per donor, reduced to 30 principal components and passed to logistic regression. Under the same cross-validation it separates responders from non-responders at an AUROC of 0.93 [95% CI 0.84–0.99] (Figure 3f). The response label is thus nearly recoverable from the molecular endpoint by construction — a useful property when clinical endpoints are unavailable, as in organoids, cell lines and mouse models.

The gap between the prospective and retrospective tasks is large. Whether it reflects the model, the sample size, or unmeasured confounders and donor-specific dynamics that make each transition individualized is an open question; the benchmarking below provides some insights.

We also tested the standard alternative of training the latents end-to-end on the endpoints. A symmetric multi-task model degrades state prediction while leaving response near chance at this sample size, and letting the state objective alone reshape the representation did not beat freezing it: finite per-neighborhood supervision starves the representation relative to the abundant per-cell signal (Supplementary Note 1). One representation with many decoupled readers is the architectural finding of this work, and it is what keeps the backward analysis stable, since gene attribution never inherits the noise of the response classifier.

## *Benchmarking against existing methods*

We benchmarked state-change prediction against scGen latent vector arithmetic [7], CellFlow [21,22] on both gene-expression and Geneformer-V2 inputs, and scDEFT without neighborhood aggregation, under matched donor-grouped cross-validation (Table 1; Methods). scGen reached 18% of the reproducibility headroom and CellFlow did not exceed the baseline, perhaps indicating that the single-perturbation, donor-held-out task does not exercise its large-scale cross-perturbation design. Removing neighborhood aggregation from scDEFT dropped it from 45% to 14% of headroom, identifying the role of aggregation in extracting signal from a noisy background.

**Table 1 |** State-change prediction benchmarking**.** *scDEFT's neighborhood state-change head versus standard perturbation-prediction baselines on the IBD atlas under matched donor-grouped cross-validation (see Methods). Metric: cosine between predicted and observed pre- to post-treatment shift, measured per donor and neighborhood and averaged. Also, reported as a percentage of the reproducibility headroom between the cohort-average-shift baseline and the split-half reproducibility ceiling. *scGen-VAE uses its own scoring in HVG gene space, and the code is our implementation of Lotfollahi 2019 algorithm [7].*

| Method | Cosine of pred-shift and actual-shift, cross-validated | | Baseline | Ceiling | % of the headroom |
|---|---|---|---|---|---|
| | Mean (SD) | 95% CI | | | |
| scGen VAE* (latent arithmetic) | 0.24 (0.01) | [0.23, 0.25] | 0.08 | 0.83 | 20% |
| CellFlow, gene-expr, PCA 100 | 0.02 (0.12) | [-0.05, 0.09] | 0.05 | 0.88 | -4% |
| CellFlow, Geneformer V2 emb. | 0.03 (0.04) | [ 0.01, 0.05] | 0.08 | 0.83 | -8% |
| scDEFT**,** w/o neighborhoods | 0.23 (0.04) | [0.20, 0.26] | 0.15 | 0.72 | 14% |
| scDEFT**,** neighborhoods, MIL-attn | **0.38** (0.07) | [0.33, 0.43] | 0.15 | 0.72 | **45%** |

**Table 2 |** Responder-prediction benchmarking**.** *Prospective (from pre-treatment cells only) and retrospective (includes post-treatment cells) responder/non-responder prediction versus standard baselines on the IBD atlas under identical, response-stratified 10-fold cross-validation. Metric: AUROC across the folds.*

| Method | AUROC, cross-validation mean [95% CI] | |
|---|---|---|
| | Prospective Prediction | Retrospective Classification |
| Gene Signature - GIMATS | 0.46 [0.29, 0.63] | 0.81 [0.68, 0.93] |
| Gene Signature - OSM | 0.49 [0.33, 0.64] | 0.64 [0.49, 0.78] |
| Pseudo-bulk (gene space) -> PCA -> LR | 0.50 [0.33, 0.66] | **0.95** [0.87, 0.99] |
| Pseudo-bulk (gene space) -> RF (400) | 0.48 [0.32, 0.65] | 0.91 [0.80, 0.99] |
| Composition (K=80) -> LR | 0.57 [0.39, 0.73] | 0.87 [0.75, 0.95] |
| scDEFT, latents, w/o neighborhoods | 0.55 [0.39, 0.71] | N/A |
| scDEFT, latents, with neighborhoods | **0.70** [0.56, 0.84] | N/A |
| scDEFT, transition post - pre, by nbhds | N/A | 0.74 [0.59, 0.87] |
| scDEFT, transition post - pre, by cell emb. | N/A | 0.80 [0.67, 0.91] |
| scDEFT, post-cell embeddings mean pool | N/A | 0.93 [0.84, 0.99] |

Responder prediction was benchmarked against published gene-set signatures (GIMATS [10], OSM [11]), pseudobulk logistic-regression and random-forest classifiers, and a neighborhood-composition classifier, under the same cross-validation (Table 2; Methods).

Prospectively, scDEFT without neighborhood aggregation reached 0.55 AUROC, barely above chance; with neighborhoods it reached 0.70 [95% CI 0.56–0.84], the only method whose interval excludes chance. Every comparator's interval includes 0.5, indicating the challenge of predicting response with a small number of donors.

Retrospectively the task is far easier: pseudobulk logistic regression reaches 0.95 AUROC and scDEFT's post-treatment classifier 0.93, confirming that the label is almost fully recoverable from the endpoint.

Most recoverable signal therefore sits in the treated state, as expected, yet scDEFT's latents extract usable prospective signal, and a simple transition-based model at 0.80 AUROC indicates remaining headroom in learning the transition itself.

## Gene-driver attribution of drug response

The forward heads are drug-conditioned but do not, by themselves, identify the genes or programs behind a prediction. Linear models expose feature weights directly; deep models do not. scDEFT therefore adds a backward attribution stage that traverses from the outcome back to the genes whose expression drives, and is affected by, drug response.

The backward stage operates one compartment at a time on pre-treatment cells. Their latents are mean-pooled separately for responders and non-responders and differenced per dimension; the 784 dimensions are ranked by the magnitude of that difference, sign-agnostically, and the ten highest-ranked survivors are retained after dimensions dominated by housekeeping, translation or stress genes are dropped. Each retained driver is attributed to genes by Spearman correlation with measured expression, screened by a variance decomposition that separates within-cell programs from cell-composition axes, and named by pathway enrichment (Methods). Correlation was chosen over gradient- and perturbation-based attribution because it is model-agnostic and scales to hundreds of thousands of cells; its one weakness, conflating a program with composition, is handled with variance decomposition (Supplementary Note 2).

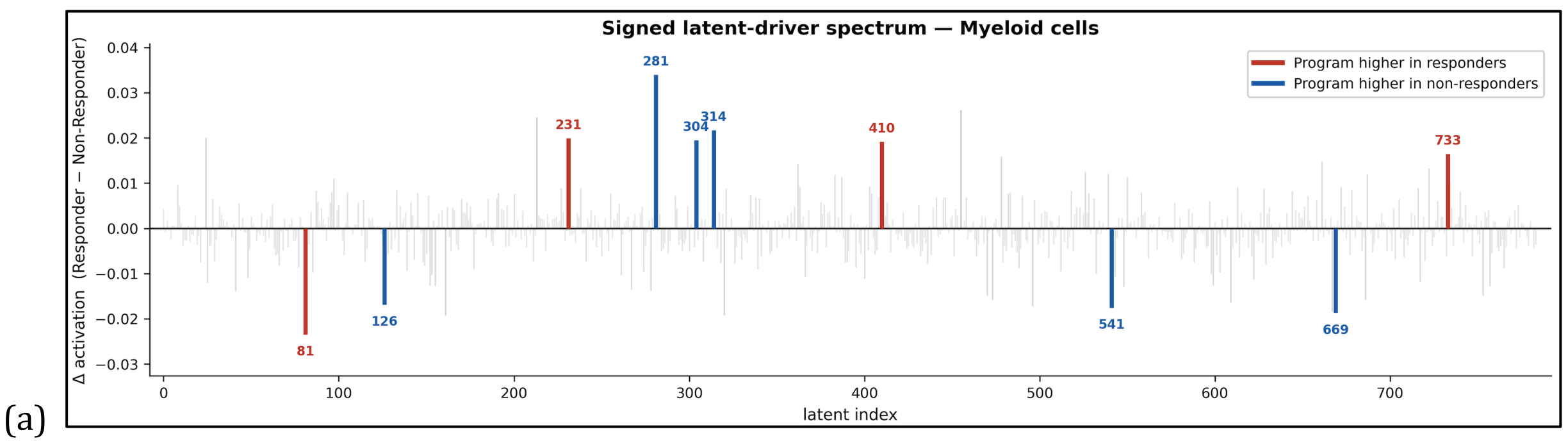

(a)

| Program | Driver | Higher In | Enriched terms | Adj. P | Top correlated genes |
|---|---|---|---|---|---|
| Inflammatory TNF-NF-kB / chemokine | 281 | Non-responder | TNF-alpha Signaling via NF-kB | 2.2e-55 | S100A9; S100A8; CCL3; SOD2; CCL4; DUSP6; MAFB; SMIM25; C5AR1; SERPINA1 |
| | | | Inflammatory Response | 3.6e-27 | |
| | | | Interferon Gamma Response | 1.6e-20 | |
| MHC class-II antigen presentation | 314 | Non-responder | Peptide Antigen Assembly With MHC Protein Complex (GO:0002501) | 4.5e-09 | CD74; HLA-DQB1; ARL4C; RAB11FIP1; HLA-DQA1; CIITA; HLA-DRB1; CXCR4; CCDC88A; HLA-DMB |
| | | | MHC Class II Protein Complex Assembly (GO:0002399) | 1.3e-08 | |
| | | | Peptide Antigen Assembly With MHC Class II Protein Complex (GO:0002503) | 1.3e-08 | |
| TREM2-DAP12 lysosomal macrophage | 231 | Responder | Complement | 5.7e-13 | FCER1G; TYROBP; PSAP; ASAH1; CTSS; GLUL; SH3BGRL3; CTSB; S100A4; LAPTM5 |
| | | | Vacuolar Acidification (GO:0007035) | 1e-05 | |
| | | | Microglial Cell Activation (GO:0001774) | 1.4e-05 | |

(b)

**Figure 4 | Backward attribution for Myeloid cells**. *Anti-TNF cohort, 14,875 pre-treatment myeloid cells. (a) Δ activation (mean responder − mean non-responder) for all 784 latent dimensions; the ten retained drivers are emphasized. Color denotes the direction of the gene program the dimension carries — assigned by sign(Δ) × sign(ρ) — which inverts when genes correlate negatively with the dimension. (b) The three top programs, each with its lead driver, enriched terms (Enrichr, BH-adjusted P) and most strongly correlated genes. All drivers shown passed the variance-decomposition control as within-cell programs (see the text).*

The backward attribution was run on the Thomas anti-TNF cohort [2], one compartment at a time, using the model trained across all three cohorts and using pre-treatment cells only (Figure 4). We report the two compartments carrying the dominant anti-TNF response axis whose drivers resolved as within-cell programs under that control: the epithelium (colonic and ileal together) and the myeloid compartment. All ten epithelial drivers scored as programs; nine of ten myeloid drivers did.

Epithelial compartment. Three programs dominate the baseline separation: an antimicrobial and secretory barrier program elevated in future responders, carrying the largest driver in either compartment (responder-minus-non-responder gap 0.043; LCN2, PLA2G2A, DUOX2, DMBT1, PIGR, MUC2, MUC5B); a cell-cycle program elevated in non-responders (MKI67, TOP2A, CDK1, BIRC5), whose E2F and G2–M enrichment is the most extreme in the analysis; and a mature absorptive-colonocyte program, also on the non-responder side (FABP1, CA1, CA2, LGALS4), carrying a strong oxidative-phosphorylation signature. A stem and regenerative axis (ASCL2, RGMB, EPHB3, SMOC2) sits on the responder side of the same dimensions. All ten drivers scored as within-cell programs rather than composition artifacts (between-subtype variance ratio ≤ 0.14; retained within-subtype gap ≥ 0.82).

Myeloid compartment. Three programs dominate. An inflammatory TNF–NF-κB program elevated in future non-responders carries the largest driver here (gap 0.034, retained 0.74; S100A8, S100A9, CCL3, CCL4, SOD2), led by TNF-α signaling via NF-κB well ahead of any other term. MHC class-II antigen presentation, also on the non-responder side, is the most reproducible result in this compartment: three independent drivers, localizing to different myeloid states, each place CD74, HLA-DQA1, HLA-DQB1 and CIITA on that arm. A repair-and-regulatory macrophage program is elevated in responders, led by a TREM2/DAP12 lysosomal axis (TYROBP, FCER1G, PSAP, CTSS, CTSB). Nine of ten drivers scored as within-cell programs; the exception, an $S100A8/9^{+}FCN1^{+}$ monocyte dimension, is reclassified by the composition control as cell identity rather than response signal (Figure 4; Supplementary Note 3).

By comparison, unguided differential expression followed by GSEA is cross-cohort reproducible but dominated by the most abundant compartment, its top program being generic T-cell-receptor/PD-1 signaling. scDEFT computes its contrast within each compartment, which lets the programs be read compartment by compartment.

These drivers recapitulate the principal cell-state associations reported for this cohort by Thomas et al. [2], reached by an independent route involving latent attribution rather than abundance and differential-expression testing. Their goblet/mucin and stem-reconstitution signatures in responders, their $TREM2^{+}$ pro-repair macrophages in remission, and their S100A8/9-high inflammatory monocytes in non-remission are each recovered. Two findings differ: the C1Q genes load to the non-responder side here, and we do not reproduce their compartment-specific interferon split, since an interferon-γ term is enriched on seven of the ten myeloid drivers and on both arms. The full concordance analysis is in Supplementary Note 3.

## Applications

scDEFT turns predictions and attributions into decision-ready outputs: a cell-type-resolved target list, a co-target nomination with a traceable evidence trail, patient stratification, and counterfactual queries. The IBD analysis is a worked example demonstrating these capabilities, not a claim to have solved IBD.

Read as target hypotheses, the two programs that separate future non-responders in the myeloid compartment converge on one upstream node. The inflammatory driver names its own receptors (S100A8/S100A9 through TLR4 and RAGE, CCL3/CCL4 through CCR1 and CCR5) while the antigen-presentation driver is headed by CIITA, whose interferon-responsive promoter is driven by STAT1 and IRF1, placing JAK1/JAK2 above both: interferon-γ → JAK–STAT1 → IRF1 → CIITA → MHC class II. The same co-target is reached independently from the pre-to-post trajectory, where a lasting interferon and JAK–STAT program is the process anti-TNF fails to resolve in non-responders (interferon-γ and interferon-α responses, $P = 9.1 \times 10^{-32}$ and $3.2 \times 10^{-24}$), and it resolves on treatment in held-out JAK-inhibitor responders. The method therefore nominates the add-on prospectively and then stress-tests it on a cohort that received that drug. It is a concrete, testable hypothesis rather than a proven therapy (Supplementary Note 4).

Because response is predicted from pre-treatment cells alone, patients can be sorted before drug treatment to enrich a trial cohort, to weigh whether to start a drug, or to rescue a drug that missed its endpoint in an unselected population by identifying the subgroup in which it works. While on the three cohorts the prospective head reaches 0.70 AUROC, confirmation in patients outside these cohorts remains future work.

Because the representation is trained to reproduce each per-cell transition, a drug can be applied as a transport operator to cells it was never observed on. Applying the tofacitinib operator to ulcerative-colitis cells from the anti-TNF cohort predicts a shift more concordant with the observed tofacitinib effect than the wrong-drug adalimumab operator (cosine 0.289 vs 0.185, against a ceiling of 0.72 and a baseline of 0.15), and the ordering holds neighborhood by neighborhood (86% of 56 neighborhoods, which are present in both cohorts; one-sided Wilcoxon $P = 1.7 \times 10^{-9}$). This is a proof of concept for counterfactual queries (Figure 5 and Table 3).

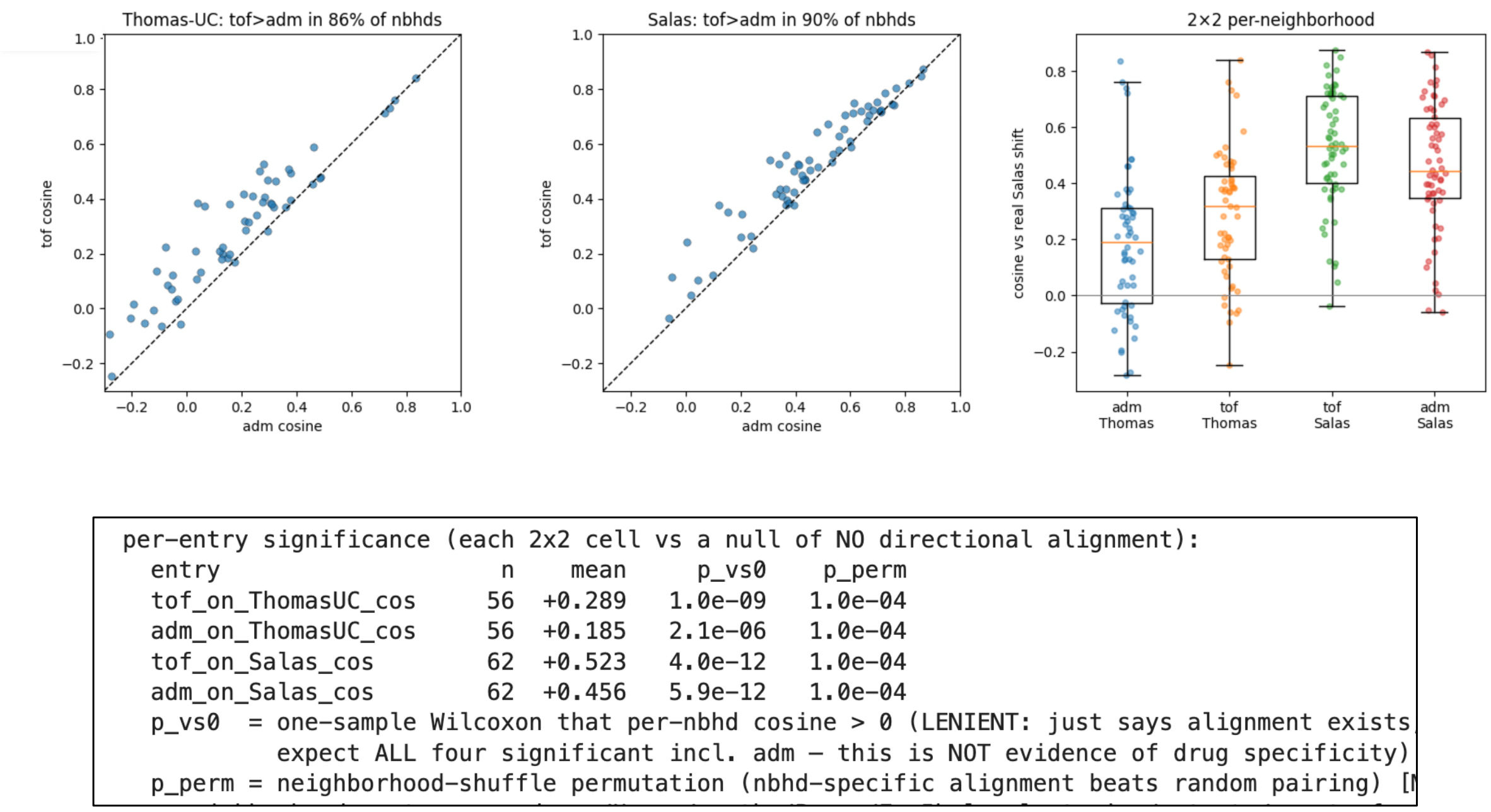

**Figure 5 | Counterfactual drug transport, per neighborhood**. *Directional agreement (cosine) with the observed tofacitinib shift, neighborhood by neighborhood, for the tofacitinib operator versus the adalimumab operator applied to the same Thomas-UC cells. Tofacitinib is more concordant in 86% of the 56 neighborhoods (paired mean difference +0.104; one-sided Wilcoxon signed-rank $P = 1.7 \times 10^{-9}$).*

**Table 3 |** *Counterfactual drug transport across cohorts. Directional agreement (cosine similarity) between each predicted per-neighborhood shift and the observed post-treatment shift of the Salas tofacitinib cohort. Rows give the drug whose learned operator was applied; columns give the cohort whose pre-treatment cells it was applied to. Tofacitinib on Salas cells (+0.523) is the native case the model was trained to predict; tofacitinib on Thomas-UC cells (+0.289) is the counterfactual of interest; the adalimumab row is the wrong-drug control. The final row reports a different quantity — the agreement between the two tofacitinib predictions themselves, i.e. consistency of the operator across cohorts, not agreement with the observed shift. For scale, the split-half reproducibility ceiling is ≈0.72 and the shared-drug baseline ≈0.15.*

| Drug applied | Cohort whose pre-treatment cells the operator was applied to | |
|---|---|---|
| | Thomas-UC (rel. to pred. Salas/tof) | Salas (rel. to pred. Salas/tof) |
| Tofacitinib | **+0.289 (55%)** | +0.523 (100%) |
| Adalimumab | +0.185 (35%) | +0.456 (87%) |
| Operator consistency (a different comparison): cosine similarity between the two tofacitinib predictions themselves — Pred Thomas-UC vs Pred Salas = +0.473 | | |

## Discussion

scDEFT reframes longitudinal single-cell analysis of drug treatment as forward prediction followed by backward analysis. Three design choices make it work. The transcriptional neighborhood is fine enough to localize a program and coarse enough to be reproducible, which lifts state-change prediction to nearly half its reproducibility ceiling. Decoupling beats coupling at realistic sample sizes. And the backward stage reads its explanation out of the same representation that produced the predictions, so the mechanism reported is tied directly to the predictive signal.

The result is a method that is simultaneously predictive and explanatory, and whose explanation is actionable: it names prospectively the programs that fail to resolve in eventual non-responders and nominates co-targets that address them. Current performance is bounded mainly by data scale (51 donors, three cohorts, two drug classes) rather than by design. The framework is disease- and drug-agnostic: any longitudinal atlas with paired samples and an outcome label can be analyzed this way, and as such datasets accumulate across oncology [19] and immunology we expect the forward–backward, neighborhood-decoupled design to generalize.

Comparison with a standard differential-expression analysis of the same cells (pseudo bulk, DESeq2) [15] shows that the two agree on a pro-repair macrophage and oxidative-metabolic axis in responders but disagree on MHC class-II antigen presentation and on resident-macrophage markers, which three independent latent drivers place with non-responders. The analyses differ in what they hold constant: the test compares donor pseudobulk without a within-subtype control (Supplementary Note 5).

Perturbation-trained foundation models such as Tahoe-x1 [23, 24] learn general representations from giga-scale cell-line screens but never observe a clinical outcome or the same patient before and after therapy, so they cannot by themselves say which patients will respond. Since scDEFT representation is embedding-agnostic, such a model is a natural upstream component. On the most comparable benchmark — a held-out cellular context or donor — Tahoe-x1 paired with the State model [25] recovers a fraction of achievable signal close to scDEFT's 45% of its reproducibility ceiling, but scDEFT reaches this on primary patient tissue with real therapies, and predicts the shift with its own head tied to clinical response (Supplementary Note 6).

scDEFT operates at atlas scale: neighborhood assignment is linear in the number of cells, avoiding an explicit neighbor graph over millions, and because the latents are trained jointly across cohorts and read over a shared scaffold, with a cross-cohort sign filter on reported genes, the retained signal is the part common to independent studies rather than any single batch.

# Methods

**Atlas assembly and harmonization.** Three published longitudinal IBD single-cell cohorts — TAURUS anti-TNF (adalimumab) [2], an infliximab cohort (GSE298464) [4], and a tofacitinib cohort (GSE253006) [3] — were assembled into a single atlas of 1,156,405 cells. Donor, timepoint (Pre/Post), compartment, drug, and clinical response (R/NR) were harmonized into a common schema. Compartment is a collection of related cell types as defined in Thomas et al [2]. Fifty-one donors had paired pre/post samples with response labels and were used for modeling; 518,437 pre-treatment cells entered the prediction tasks. No batch integration was applied to the embeddings beyond the shared neighborhood scaffold; cohort of origin is retained as metadata and used in the cross-cohort gene filter.

**Cell and drug embeddings.** Each cell was embedded with zero-shot Geneformer-V2 [1] into a 768-dimensional vector. A low-dimensional context code (compartment) was concatenated to form the biological input to the representation. The context code can be extended to other metadata of cells such as inflammation score, specific disease, and other patient and disease data. Drugs are embedded with ESM3 [27] for biologics, using their sequence vectors (light chains and heavy chains), and with ChemBERTa [26] for small molecules, using their SMILES strings.

**Transcriptional neighborhoods.** A shared scaffold of K = 80 neighborhoods, which is a hyper parameter that we tuned over a range of 30 to 200, was defined by K-means clustering of cell embeddings pooled across all cohorts, so a neighborhood index denotes the same transcriptional state in every donor. All downstream targets and contrasts are computed per (donor, neighborhood).

**Shared representation.** The drug-conditioned cell representation is produced by feature-wise linear modulation (FiLM; Perez et al. [20]). Each cell's biological input is its Geneformer-V2 embedding $g$ (768-dimensional) concatenated with a 16-dimensional compartment context code, giving $e$ of dimension 784.

Each drug is a frozen 1,536-dimensional embedding $d$ from ESM3 — zero-padded to 1,536, for small molecules. A two-layer generator maps the drug to a per-feature scale $\gamma(d)$ and shift $\beta(d)$:

$$h = \mathrm{GELU}(W_1\, d + b_1)$$

$$[\gamma(d); \beta(d)] = W_2\, h + b_2$$

where $W_1$ is 784 × 1,536 and $W_2$ is 1,568 × 784 (which are trained during the model training), and $\gamma(d)$ and $\beta(d)$ are each 784-dimensional; the generator's output layer is initialized to zero, so that an untrained drug leaves the cell unchanged. The input cell representation $e$ is drug conditioned using:

$$u = (1 + \gamma(d)) \odot e + \beta(d)$$

Layer normalization of $u$ followed by a feed-forward projection trunk yields the per-cell latent:

$$\ell = \mathrm{trunk}(\mathrm{LayerNorm}(u))$$

The latent $\ell$ is also 784-dimensional. Because $\gamma$ and $\beta$ depend only on $d$, every cell treated with the same drug should receive the same modulation.

The representation is trained with a per-cell shift prediction head. From each pre-treatment cell's latent a shift predictor, a 784 → 512 → 768 multilayer perceptron, outputs a predicted drug-induced shift $\delta$ = state-predictor($\ell$) of dimension 768, optimized by the cosine loss $1 - \cos(\delta, T - g)$, where $T$ is the post-treatment mean expression of the cell's transcriptional neighborhood and $g$ the cell's Geneformer embedding, so that $T - g$ is the observed per-cell shift. Because the objective is a cosine, training matches the direction of the per-cell shift — the move toward the neighborhood mean — not the absolute post-treatment position or the Euclidean closeness of the predicted endpoint to $T$; here $g$ is the cell's own pre-treatment Geneformer-V2 embedding (768-dimensional), so $T - g$ is the observed shift and $\delta$ is rewarded for pointing along it. Neighborhoods enter training only through this target (the model input carries no neighborhood identity) and supervision is abundant, at one target per cell. The neighborhood target is pooled over training donors within each cross-validation fold, so held-out donors never inform the representation.

After this stage the FiLM generator, trunk, and the trained latents are frozen and the latent $\ell$ is recorded for every pre-treatment cell; the shift predictor serves only to shape $\ell$ and is not used thereafter. Each downstream head, the neighborhood state-change attention-MIL head and the response probe, reads the frozen latents through its own lightweight predictor.

**State-change head and metric.** The state target for a (donor, neighborhood) pair is the shift between its post- and pre-treatment mean expression. Predictions are scored by the cosine between predicted and observed shifts, grouped by (donor, neighborhood) and averaged per donor. The baseline reference is the cohort-average shift computed from training donors; the ceiling is the split-half reproducibility of the observed shift. The deliverable head pools a neighborhood's cells with a gated attention multiple-instance layer and predicts the shift. Models were trained on Apple MPS/CUDA with AdamW and the run-to-run variation was assessed across seeds and folds.

Specifically, the pooling head is a gated-attention multiple-instance layer (Ilse et al. [14]) applied to the frozen per-cell latents. For the cells $i$ in a (donor, neighborhood) bag $B$, each latent $\ell_i$ receives a gated-attention score

$$s_i = w^T[\tanh(V\, \ell_i) \odot \sigma(U\, \ell_i)]$$

where $V$ and $U$ project the latent into a 64-dimensional attention space and $w$ reduces it to a scalar — which is normalized across the bag as $a_i = \mathrm{softmax}(s_i)$; the cells are then pooled into a single neighborhood vector $z = \Sigma\, a_i\, \ell_i$, of dimension 784. A linear projection predicts that neighborhood's shift,

$\hat{\Delta} = W z + b$

of dimension 768 — the quantity scored against the observed shift above. Only the attention parameters V, U and w and the projection W, b are trained at this stage; the per-cell latents $\ell_i$ remain frozen.

**Benchmarking.** State-change comparators were scored on the same standardized scale as scDEFT but in each method's own output space. scGen-style latent arithmetic is a re-implementation of the Lotfollahi et al. [7] algorithm: a variational autoencoder over each cohort's 2,000 highly variable genes, with the pre-to-post shift estimated as a per-neighborhood latent difference on training donors and applied to held-out donors, then decoded. CellFlow was run on gene-expression and Geneformer-V2 inputs at near-default settings. For the scDEFT without neighborhoods, the frozen pre-treatment latents were read one cell at a time, the model predicted its shift, cosine similarity is calculated with the actual shift of its neighborhood mean. The resulting per-cell cosines were averaged within each (donor, neighborhood) so that scoring, baseline and ceiling were identical to the aggregated alternative. Cells were subsampled per neighborhood so that both alternatives (i.e. w/o and with neighborhoods) took the same number of optimizer steps. Responder comparators were the published GIMATS and OSM gene-set signatures scored per donor, pseudo bulk logistic-regression and random-forest classifiers over donor-mean expression, and a classifier over the K = 80 neighborhood-proportion vector.

Two properties of the gene-space comparators differ from the scDEFT protocol. First, they are cross validated within each cohort, not over the pooled donor set. Highly variable genes are selected per cohort, so the three cohorts do not share a feature basis, and a single model cannot be fit across them; scDEFT has no such constraint because its 768-dimensional embedding space is common to every cell in every cohort, which the shared representation buys. Second, the gene-space runs use five folds rather than ten, since each fold trains a separate autoencoder per cohort. Both choices favor the comparator because each model faces one cohort's manifold with no cross-cohort batch effect to absorb and so the reported gap is conservative. For every method in Table 1, Mean (SD) is taken across cross-validation folds and the 95% confidence interval is the corresponding normal-approximation interval on the fold mean.

**Response heads**. Two response heads read the frozen representation, one a prospective predictor and the other a retrospective classifier. The prospective predictor uses pre-treatment cells only. For each donor it takes the drug-conditioned latents of those cells, averages them within each neighborhood, and then averages the resulting per-neighborhood vectors into one donor vector. The donor vectors are standardized, reduced to 30 principal components, and classified by L2-regularized logistic regression. Donors are split by a response-stratified, donor-grouped ten-fold scheme, so each donor is held out exactly once and no donor appears in both training and test. The retrospective classifier instead uses post-treatment cells and the Geneformer-V2 cell embeddings. For each donor a sample of post-treatment cells (up to 3,000) is averaged into one donor vector, standardized, reduced to 30 principal components, and classified in the same way, under the same donor-grouped ten-fold scheme repeated over five random seeds. Confidence intervals come from a donor-level bootstrap.

**Backward gene attribution**. The backward stage reads the drug-conditioned representation itself, one cohort and one compartment at a time, over the pre-treatment cells of that compartment, subsampled to 150,000 cells stratified by compartment and response when larger, for the cells that carry a responder or non-responder label. For each of the 784 latent dimensions, the mean activation over responder cells and over non-responder cells is computed, and their difference, responder mean minus non-responder mean, measures how strongly the dimension separates the two outcomes. Restricting the analysis to a subset of neighborhoods was evaluated and rejected: it reduces the cell count several-fold, inflates the responder-versus-non-responder gap, and drives the top dimensions toward cell-identity axes. Activations are pooled across the cells of that set; unlike the state-change head and the per-(donor, neighborhood) pseudo-bulk check described below, the driver contrast is not stratified by neighborhood, so the scaffold localizes which cells are examined rather than how the contrast is computed. Each driver is mapped to genes by the Spearman correlation between its activation and each gene's measured expression across the same cells; every gene is assigned to the driver it tracks most closely and given a sign from that driver's responder-versus-non-responder direction, so that genes fall on a responder side or a non-responder side. Housekeeping, ribosomal, mitochondrial, and clone or long-non-coding artifact genes are masked before correlation. Each driver is also passed through a composition control. Finally, the responder-side and non-responder-side gene sets of the retained drivers are tested for pathway enrichment with Enrichr against MSigDB Hallmark, GO Biological Process, and Reactome, and FDR-significant terms are reported. The output robustness check compared responders and non-responders by per-(donor, neighborhood) pseudo bulk with Welch t-tests, Benjamini–Hochberg FDR, and a cross-cohort sign-agreement filter; pathway enrichment used Enrichr at FDR < 0.05.

**Counterfactual queries.** A drug is treated as a transition operator by taking the trained, frozen drug-conditioned model and setting its drug token to the queried drug; passing a cohort's pre-treatment cells through the model in this configuration (a forward pass, with no retraining) yields the latents that drug would induce, including for a drug the cohort has never seen. These latents are read by the neighborhood state-change head to give a predicted shift for each donor and neighborhood, averaged across donors to one shift per neighborhood. The reference is the observed pre-to-post shift, per neighborhood, in the cohort that did receive the drug. Agreement reported here is the cosine between predicted and observed per-neighborhood shifts, and drug-specificity is a paired comparison, neighborhood by neighborhood (that are present in both cohorts), of the queried-drug operator against the cohort's own-drug operator applied to the same cells.

## Acknowledgments

I thank my cofounders at Causal Data AI, Rhodemann Li and Fariedeh Moeinvaziri, PhD, for the partnership that set the context for this work. I thank Tejal Patwari, PhD, for reviewing this work from a bioinformatics perspective. I thank Rithika Devarakonda for conducting the differential gene expression analysis of myeloid cells in the Thomas et al. cohort, the results of which I summarized in the text.

**Competing interests**. The author is a cofounder of Causal Data AI. The method described here is the subject of a pending U.S. provisional patent application.

**Funding.** This work received no external funding.

## Data and code availability

The atlas was derived from public cohorts (GSE282122, GSE298464, GSE253006). scDEFT code — the neighborhood scaffold builder, the backbone, forward heads, and backward attribution pipeline — will be available after publication.

# Supplementary Information

*scDEFT: A deep learning framework for drug-effect prediction and counterfactual reasoning*

## Supplementary Note 1

### Decoupled versus end-to-end training

In deep learning, the standard approach is to train the latents using the endpoints - state change and remission status - jointly as the learning objectives and let the training update the shared latents and model weights throughout. We found that this consistently led to poor results for our data. A symmetric multi-task model, in which the response gradient reshapes the shared trunk, degrades state prediction while leaving response itself near chance at this sample size. Letting the state objective alone reshape the representation end-to-end (an asymmetric “unfreeze”) did not beat freezing it either. The finite per-neighborhood supervision seems to starve the representation relative to the abundant per-cell signal. The winning configuration is asymmetric and decoupled: the state objective trains the latents and the drug-conditioning trunk on the plentiful per-cell signal; the prediction heads read that representation frozen and operate on the neighborhood basis. Every attempt to couple the two objectives underperformed this decoupling. This is the architectural learning from scDEFT — one representation, many decoupled readers — and it is what makes the backward analysis stable, because the gene attribution never has to inherit the noise of the response classifier. This could be a limitation of the limited number of cohorts, or it is intrinsic to this task and data. It is an open research question.

## Supplementary Note 2

### Why not SHAP or LIME

Two model-explanation methods in wide use, SHAP and LIME, are not suitable for our data and model. Both explain a single model output for a single instance by attributing it to that instance’s input features. What we need is an insight into which elements of the representation separate two groups of donors (i.e. responders and non-responders): A population-level contrast defined over the frozen drug-conditioned representation, grouped by the donors’ response labels. Applied here, SHAP would have to treat each of the 784 dimensions as a separate output and attribute it across roughly 33,000 genes for every cell, and because the backbone consumes a foundation-model cell embedding rather than raw counts, that attribution must be propagated back through the embedding model as well; the cost approaches that of the counterfactual route without its causal warrant. LIME fits a local surrogate by resampling perturbed feature vectors around one cell, which on sparse count data generates profiles off the biological manifold, and yields an explanation that is local to that cell and sensitive to resampling, whereas the quantity we report is a single cohort-level statement per dimension. Correlation against measured expression answers the group-contrast question directly.

## Supplementary Note 3

### Compartment programs identified in Thomas et al. dataset

**Epithelial compartment.** scDEFT analysis shows that three programs dominate the pre-treatment separation between responders and non-responders. First, an antimicrobial and secretory barrier program elevated in future responders, combining Paneth and goblet output with the DUOX2 oxidative-defense system (LCN2, PLA2G2A, DUOX2, DUOXA2, DMBT1, PIGR, MUC1, MUC4, MUC5B, MUC12, GPX2); its enrichment is led by antigen assembly and presentation, with maintenance of gastrointestinal epithelium alongside. Second, a cell-cycle program elevated in non-responders (MKI67, TOP2A, CDK1,

UBE2C, BIRC5, PCLAF), whose enrichment is the most extreme in the whole analysis — E2F targets, G2–M checkpoint and mitotic chromosome condensation at combined scores above 10,000. Third, a mature absorptive-colonocyte program, also on the non-responder side (FABP1, CA1, CA2, SLC26A2, LGALS4, PHGR1, ADH1C, HMGCS2), which carries an unusually strong oxidative-phosphorylation signature (combined scores 25,000–27,000). A stem and regenerative axis (ASCL2, RGMB, EPHB3, SMOC2) sits on the responder side of the same dimensions that carry the mature-colonocyte genes on the non-responder side. All ten drivers scored as within-cell programs rather than composition artifacts (between-subtype variance ratio ≤ 0.14; within-subtype responder gap retained ≥ 0.82). Cell-state localization is indicative rather than exclusive: Paneth-labeled dimensions appear on both arms, carrying secretory genes toward responders and absorptive genes toward non-responders. Unlike the myeloid compartment, where oxidative phosphorylation loads to the responder side, here it resolves clearly to the non-responder side; we therefore read the metabolic axis as compartment-specific rather than absent.

**Myeloid compartment.** Three programs dominate. First, and the largest single driver here (gap 0.034, retained gap 0.74), an inflammatory TNF–NF-κB program elevated in future non-responders (S100A8, S100A9, CCL3, CCL4, SOD2, MAFB), whose enrichment is led by TNF-α signaling via NF-κB well ahead of any other term, with leukocyte aggregation, neutrophil chemotaxis, complement-receptor signaling and an interferon-γ response term following. Second, MHC class-II antigen presentation, also on the non-responder side, and the most reproducible result in this compartment: three independent drivers, localizing to different myeloid states, each place CD74, HLA-DQA1, HLA-DQB1, HLA-DPB1 and CIITA on that arm with MHC class-II assembly as the top enriched term. Third, a repair-and-regulatory macrophage program elevated in responders, with three facets — a TREM2/DAP12 lysosomal axis (TYROBP, FCER1G, PSAP, CTSS, ASAH1, CTSB; complement, vacuolar-acidification and microglial-activation terms), an immune-dampening axis (PSAP, CD44, ZEB2, CYBB; immune-response-inhibiting receptor signaling and negative regulation of leukocyte proliferation), and an oxidative-phosphorylation axis built from nuclear-encoded electron-transport genes. Dendritic-cell drivers cDC2 (FCER1A, CD1C, CLEC10A, JAML) and cDC1 (CLEC9A, DNASE1L3, CST3), and an iron-handling resident-macrophage axis (SLC40A1, C1QA, C1QC, FCGRT, SELENOP) also load to the non-responder side; a mast-cell axis (MS4A2, HPGDS) loads to the responder side. Nine of ten drivers scored as within-cell programs; the exception is an S100A8/9⁺FCN1⁺ monocyte dimension that the composition control flags as mixed (between-subtype variance ratio 0.18, retained gap 0.53), tracking which cells are monocytes more than what those monocytes are doing.

## Concordance with Thomas et al. study

**Concordant on the major axes.** The compartment-resolved drivers recapitulate the principal cell-state associations Thomas et al. [2] report for this cohort, reached by an AI-based drug-conditioned latent attribution approach rather than through cell-state abundance and differential-expression testing. In the epithelium the prior study found that responders enter treatment with a more differentiated, better-defended epithelium — goblet/mucin enrichment (MUC2, MUC5B; their Fig. 4g,h) and LGR5⁺ stem/epithelial reconstitution in remission — each mirrored here by a responder-elevated driver: the goblet-secretory program, the Paneth antimicrobial program, and the ASCL2/RGMB/EPHB3/SMOC2 stem-regenerative axis. In myeloid they identify TREM2⁺ C1Q-high IL1B-low pro-repair macrophages in CD remission (their Fig. 4d); our TREM2/DAP12 lysosomal-macrophage driver recovers that axis on the responder side, though the complement C1Q genes themselves load to the non-responder side here, on an iron-handling resident-macrophage dimension. So, we reproduce the TREM2 half of their signature and invert the C1Q half. The prior study also reports S100A8/9-high TNF-high inflammatory monocytes in non-remission, and the baseline-restricted analysis does recover that pole: the leading myeloid driver is an S100A8/9, CCL3/CCL4, TNF–NF-κB inflammatory program on the non-responder side. A separate S100A8/9⁺FCN1⁺ dimension scores as mixed under the composition control and sits on the responder side; it tracks monocyte identity rather than monocyte activation, which is consistent with their own observation that S100A8/9-high monocyte abundance does not differ at baseline. The two dimensions are therefore not in conflict — one carries the inflammatory program, the other the cell type that usually carries it.

**Differing in detail.** Thomas et al. resolve interferon as compartment-specific with epithelial interferon tracking remission, myeloid interferon tracking non-remission. An interferon-γ response term is enriched on the leading non-responder driver, the inflammatory TNF–NF-κB dimension, which is the direction they report — but the same term is enriched on seven of the ten myeloid drivers and on both arms, so we do not read it as an arm-specific interferon signal. The gene-set probe reaches the same conclusion from the other side: interferon programs are present in the representation but carry almost no responder separation at baseline. Individual interferon genes do not rank among the top correlates of any driver in either compartment. We therefore do not reproduce their compartment-specific interferon split in either direction. The neighboring antigen-presentation axis shows the compartment-specific split more sharply: MHC class-II (CD74, HLA-DQA1, HLA-DQB1) loads toward responders in the epithelium and toward non-responders in the myeloid compartment, the latter on three independent drivers.

**Unresolved or missing on our side.** Two of the Thomas et al. observations are not recovered by the latent readout: their γδ-T and T-cell-aggregate associations are not reproduced with a consistent direction by our T-cell drivers; and their non-remission epithelial-damage program does not appear, our non-responder epithelial drivers forming a mature absorptive-colonocyte and proliferation program instead. Their non-remission inflammatory-monocyte program, listed as missing in an earlier version of this analysis, is recovered once the gene correlations are restricted to baseline cells.

**What our analysis adds.** While the Thomas et al. study reports which cell states differ in abundance, ours attaches to each state the within-cell program it carries, quantified against a composition control (between-subtype variance ratio and retained within-subtype gap) and ranked by effect on the drug-conditioned representation, and it is this control that lets us call the S100A8/9 axis a cell-identity dimension rather than the response signal itself.

Two caveats regarding the agreement: we pool Crohn's disease and ulcerative colitis, whereas several of the effects reported by Thomas et al. were disease-specific (e.g., baseline epithelial frequency in Crohn's, not ulcerative colitis). Our results are also influenced by the analysis choices we made in the attribution analysis (such as the subset of cells analyzed, ranking, exclusions, and composition control). Despite these differences, the generally broad and strong agreement is encouraging.

# Supplementary Note 4

## Targets, co-targets and the evidence trail

Targets and Co-targets: scDEFT turns the drivers from the gene attribution into a list of candidate targets that is grounded in the patient data. These drivers are computed on pre-treatment cells alone, so that the insights are established prospectively. For each compartment, it reports the biological processes that most clearly separate patients who respond to the drug from those who do not and ranks them by how strongly they separate the two groups. The processes elevated in future responders, but not in future non-responders are precisely the ones the current treatment leaves unaddressed. Those unaddressed processes are where one would look for a new drug or an add-on. And because each process is linked to the specific cell type and genes behind it, the insights are specific.

**Reading the same baseline drivers as target hypotheses.** Tracing the two top myeloid cell drivers through the attribution analysis shows that they both converge on a single upstream control node: JAK1/JAK2 (following the pathway interferon-γ → JAK–STAT1 → IRF1 → CIITA → MHC class II). The methodology highlights that this master switch can be identified using pre-treatment cells alone and exposes alternative treatment strategies like CD40–CD40L blockade (tested via ravagalimab) and calprotectin sensing blockade (TLR4/RAGE). The model uncovers crucial cautions: direct suppression of antigen presentation would be broadly immunosuppressive since it reflects interferon tone rather than being an independent driver, and the same antigen-presentation axis acts appears protective in the epithelium but pathogenic in the myeloid compartments — a conflict unexposed by pooled analyses.

Finally, the model suggests epithelial drivers are better suited for patient stratification than as drug targets, where non-responders can be identified via routine MKI67 or TOP2A biopsy staining, while elevated effectors like DUOX2, LCN2, PLA2G2A, and PIGR in responders represent protective programs to preserve.

**Gene-expression space analysis enabled by neighborhoods.** In addition to the latents-driven state-change modeling, the neighborhoods can also help to analyze the data in other ways. As an example, the gene-expression difference between responders and non-responders within the scDEFT-nominated neighborhoods, in the anti-TNF cohorts, isolates a distinct inflammatory program in the myeloid cells that the drug fails to resolve in non-responders. It is characterized by interferon and JAK–STAT signaling (interferon-γ and interferon-α responses among the most significant, $P = 9.1 \times 10^{-32}$ and $3.2 \times 10^{-24}$). This program is seen in the pre-to-post trajectory rather than in the pre-treatment driver ranking because it diverges only under treatment. In the held-out JAK-inhibitor cohort, the program resolved on treatment and resolved more in responders than in non-responders on the same drug — a difference pharmacology alone does not explain. That makes the co-target a testable hypothesis with data-level support. It is not sufficient on its own: Salas et al. attribute non-response to myeloid hyperactivation, driven partly by loss of IL-10's anti-inflammatory signaling, which runs through the same JAK pathway the drug inhibits [3]. The co-target is therefore a hypothesis to test, not a conclusion.

# Supplementary Note 5

## Comparison with differential expression

To gauge what scDEFT adds over a standard analysis, we compared its drivers with a differential gene expression analysis, using scanpy [15], between responders and non-responders on the Thomas cohort's pre-treatment myeloid cells (pseudobulk, DESeq2). The results indicate partial agreement. Both place a pro-repair macrophage program and an oxidative-metabolic response on the responder side — the test through CD163, FABP4, HMOX1 and MT1M, the latent readout through the TREM2/DAP12 lysosomal axis and a nuclear-encoded electron-transport axis. They disagree on two programs. DGE puts MHC class-II antigen presentation (HLA-DQA2, IRF4) with responders, whereas three independent latent drivers place it with non-responders; and the DGE's resident-macrophage markers sit with responders, whereas the latent iron-handling resident-macrophage dimension (SLC40A1, C1QA, FCGRT) sits with non-responders. The two analyses differ in what they hold constant. DGE compares donor pseudo-bulk without a within-subtype control, while the latent readout keeps only the part of the gap that survives inside a cell subtype. Which of the two is right for antigen presentation requires further studies.

**The comparison also exposes what DGE cannot do.** In our contrast on pre-treatment myeloid cells, the standard differential gene expression analysis returned only a handful of genes elevated in responders against several thousand elevated in non-responders at uniform, modest fold change, which is likely a technical artifact of global composition or library-size skew across a small number of donors rather than true biology, requiring tedious manual filtering. scDEFT bypasses this entirely: its composition control retains only responder-non-responder differences surviving within local cell subtypes, while its cross-cohort sign filter discards technical and cohort-specific noise. Consequently, scDEFT reaches the same headline biology as painstakingly decontaminated DGE automatically, tying those gene programs directly to its own predictive response models.

# Supplementary Note 6

## Comparison with Tahoe-x1

Emerging perturbation-trained single-cell foundation models, such as Tahoe-x1 (Tx-1) [23], are another relevant approach to understanding drug effects. Tx-1 learns general representations of genes, cells and compounds from giga-scale small molecule compound screens on cell lines (the Tahoe-100M compendium [24] that comprises over 100 million cells from 50 cancer cell lines exposed to more than 1,100 small molecules). Such models are powerful on molecular tasks, predicting a compound's transcriptional effect, gene essentiality, or cell identity, but they are trained on immortalized cell lines perturbed in vitro, and they never observe the two things a treatment decision turns on: a clinical outcome, and the same patient before and after therapy. Their data are cross-sectional snapshots without response or resistance labels and without longitudinal structure, so a cell-line model cannot, on its own, say which patients will respond or why. scDEFT is built for that question. It is trained from longitudinal, primary patient data rather than cell lines, so both its predictions and the genes it nominates are tied to who improved and who did not. The two approaches are complementary rather than competing: because scDEFT's representation is embedding-agnostic, a foundation model of this kind is a natural upstream component, and in our own perturbation-transfer experiments the conclusions were unchanged whether cells were embedded with Geneformer-V2 or with Tahoe-x1 (70M model) — indicating that scDEFT's clinical traction comes from its longitudinal, outcome-anchored design rather than from any single embedding.

Our state-change prediction can also be read against the perturbation-response benchmarks reported for these foundation models. Notably, Tahoe-x1 does not predict a perturbation's effect on its own; it pairs its embeddings with a separate transition model, the State model (Adduri et al.) [25] and scores the result as the correlation between the predicted and observed mean expression change. In the most comparable setting, predicting the response of a held-out cellular context or donor never seen in training, that pipeline recovers a fraction of the signal achievable above simple averaging baselines. It is close to scDEFT accuracy in predicting the treatment shift for held-out patients. scDEFT reaches this in a harder regime, on primary patient tissue, real therapies, and matched pre- and post-treatment samples rather than controlled cell-line perturbations. Two design differences accompany the parity. First, the foundation-model embedding contributes little to the benchmark itself: in the held-out setting the transition model performs nearly as well on the raw gene-expression space as on the Tahoe-x1 embedding, whereas in scDEFT the neighborhood aggregation is what lifts prediction from a small fraction of the ceiling to nearly half. Second, scDEFT predicts the shift with its own head and ties it to clinical response, rather than delegating prediction to an external module trained on unlabeled perturbation screens. We therefore view the two as complementary: a foundation model of in vitro perturbation and a clinically grounded, longitudinal model of in vivo treatment response, reaching comparable accuracy on their respective hardest tasks.